%% file: Optimization.tex
\documentclass[preprint,3p,times]{elsarticle}
\input{Declarations.tex}

\journal{Mathematics and Mechanics of Solids}

\begin{document} 


\title{Maximum velocity of self-propulsion for an active segment }

\author[lms,curie]{P. Recho}
\author[lms]{L. Truskinovsky}
\ead{trusk@lms.polytechnique.fr}
\address[lms]{LMS,  CNRS-UMR  7649,
Ecole Polytechnique, Route de Saclay, 91128 Palaiseau,  France
}
\address[curie]{Physicochimie Curie, CNRS-UMR168, Institut Curie, Centre de Recherche, 26 rue d'Ulm F-75248 Paris Cedex 05, France}

\begin{abstract}
The  motor part of a crawling eukaryotic cell can be represented schematically as an active continuum layer.  The main active processes in this layer are  protrusion, originating from non-equilibrium polymerization of actin fibers,  contraction, induced by myosin molecular motors and  attachment due to  active bonding of trans-membrane proteins to a substrate. All three active mechanisms are regulated by complex signaling pathways involving chemical and mechanical feedback loops whose microscopic functioning is still poorly understood. In this situation, it is instructive to take a reverse engineering approach and study a  problem of finding the spatial organization of standard active elements inside a crawling layer ensuring  an optimal cost-performance trade-off.   In this paper  we  assume   that (in the range of interest) the energetic cost of self-propulsion is velocity independent  and adopt,  as an optimality criterion, the maximization of the overall velocity.   We then choose a prototypical setting, formulate the corresponding  variational problem and obtain a set of bounds suggesting that radically different  spatial distributions of  adhesive complexes  would be optimal depending on the  domineering active mechanism of self-propulsion.  Thus, for  contraction-dominated motility, adhesion has to cooperate with 'pullers' which localize at the trailing edge of the cell,  while  for  protrusion-dominated motility  it  must conspire with 'pushers' concentrating at the leading edge of the cell. Both types of crawling mechanisms were observed experimentally. 
\end{abstract}

\maketitle

\section{Introduction}
 
Eukariotic cells are spatially extended  active  bodies  that can steadily self-propel in  viscous environments at low Reynolds numbers  \cite{childress1981, Lauga2009}.  It has been  understood \cite{Taylor1951, Purcell1977, Desimone2008} that in these conditions a combination of stationarity  and linearity of friction leads to kinematic reversibility and that a symmetric  (under time reversal)  stroke  cannot produce self-propulsion.  For Stokes swimmers,  a variety of non-symmetric motility strategies  have been proposed and optimized using various efficiency criteria  \cite{Shapere1987, Najafi2004, Leshansky2007, Alouges2009, Alouges2011, Osterman2011, Michelin2012};  similar models for crawlers advancing on a frictional background were considered in \cite{Desimone2012, Noselli2014, Gidoni2014}.  Most of the self-propulsion  mechanisms proposed in these papers   are  fully kinematic  in the sense that  the time dependence of the shape of a swimmer/crawler is  prescribed.  This  implies that  appropriately chosen  actuators can always perform the required internal movements.  In cells the role of such actuators is played by  \emph{active agents} and in this paper we focus on the fact that their dynamics, while being driven biochemically, must be compatible with the fundamental balances of mass and  momentum.  
 
In the context of eukaryotic cell motility, a prototypical scheme of self-propulsion includes  protrusion through polymerization of actin filaments which is accompanied by dynamic assembly of focal adhesions;   myosin-driven contraction of the actin network which allows the motor part to advance a cargo,  and, finally,   detachment of adhesive contacts with a simultaneous depolymerization of actin fibers \cite{Mogilner2009}.   It is usually assumed that active polymerization ensuring protrusion can be described  as the work of spatially distributed  \emph{pushers}, generating positive   force couples,  while  active contraction can be viewed as an outcome of the  mechanical action of distributed \emph{pullers},  responsible for negative  force couples  \cite{Simha2002, Carlsson2011, Saintillan2012, Marchetti2012, Recho2013}.  The  role of ATP in reversible adhesion of adhesive patches (focal adhesions) is understood rather poorly and they are usually treated as  passive viscous binders  whose spatial distribution  may be regulated  actively \cite{Gao2011} . 
 
Since our knowledge of the mechanism controlling the transport and the intensity of  active agents performing protrusion, contraction and adhesion is rather limited,   we adopt in this paper  a semi-kinematic approach  and treat the corresponding distributions  as functional  \emph{control parameters}  constrained by the fundamental mechanical balances.  We then  pose a variational problem of finding the optimal  temporal and spatial distributions of these parameters  inside a crawling continuum body. In view of some successful attempts to  justify such reverse engineering approach \cite{Recho2014},  we  anticipate that our optimal solutions will be eventually backed by  an appropriate constitutive theory.

While the real organisms are expected to optimize some measure of a trade-off between the velocity of self-propulsion and the corresponding energy expenditure,  in this paper we make a simplifying assumption that the energetic cost of   self-propulsion is fixed  and use as optimality criterion  the maximization of  the overall velocity. We are  interested in steady translocation  and  assume that the internal distributions of mechanical parameters are compatible with the  traveling wave ansatz. This simplifying  assumption  allows us to replace the  optimization of the crawling stroke in space and time by a purely spatial optimization of the internal distribution of  active elements in the co-moving coordinate system. In the interest of analytic transparency we  use  the simplest  1D model  of actively contracting  continuum  (subjected to  viscous frictional forces)  that has been used repeatedly in the cell motility studies \cite{Abercrombie1980, Dimilla1991, Stossel1993, Ridley2003, Vicente-Manzanares2005, Kruse2006, Hoffman2009, Mogilner2009, Recho2013a}. In a setting similar to ours the dependence of cell velocity on the distribution of  active stresses   and adhesion properties  was studied in \cite{Carlsson2011} where both contraction and protrusion were represented by active couples.  Given that protrusion is usually localized at the leading edge of the cell,  we model the effect of active polymerization differently by using  Stefan type  boundary conditions  on the  edges of a crawling segment that fix the influx and the outflow of actin,  see also \cite{Julicher2007, Rubinstein2009, Recho2013}.  For the given strength of protrusion, we  prescribe the average level of contractile activity, and then search for  the optimal internal distribution of contractile and adhesive units.

Our analysis of the ensuing variational problem demonstrates that radically different distributions of focal adhesions are optimal depending on the  domineering active mechanism of self-propulsion.  Thus, for  contraction-dominated  motility, focal adhesions have to cooperate with pullers which end up localizing at the trailing edge of the cell  while  for protrusion-dominated motility they must conspire with pushers which concentrate in our model at the leading edge of the cell. Both types of crawling mechanisms have been observed  experimentally. 

The paper is organized as follows. In Section 2 we formulate the model. A simple analytically tractable case of contraction dominated motility is treated in Section 3. In Section 4 we obtain  analytically upper and low bounds for the self-propulsion  velocity in the general case. The same optimization problem was studied numerically in Section 5.  In Section 6 we  present some evidence that our  lower bound may in fact coincide with the optimal solution. The last Section 7 contains the discussion of our results.

\section{The model}

Following \cite{Kruse2006, Julicher2007}, we consider a one-dimensional segment of viscous active gel representing the cell lamellipodium on a frictional substrate. The segment  has two free boundaries which we identify as the trailing edge $l_-(t)$ and  the leading edge $l_+(t)$.  The force balance  can be written in  the form
\begin{equation}\label{chapIeq}
\partial_x\sigma=\xi v 
\end{equation}
where   $\sigma(x,t)$ is the stress and $v(x,t)$ is the  velocity. We assume that  the frictional coefficient mimicking the distribution of focal adhesions \cite{Rubinstein2009, Larripa2006, Julicher2007, Shao2010, Doubrovinski2011, Hawkins2011} is space and time dependent $\xi(x,t)\geq 0$.  The constitutive behavior of the gel is modeled by the equation \cite{Kruse2006, Julicher2007}
\begin{equation}\label{chapIeq1}
\sigma=\eta \partial_xv+\chi,
\end{equation}
where the active pre-stress $\chi(x,t)\geq 0$ accounting for the presence of myosin  molecular motors \cite{Salbreux2007, Bois2011, George2012, Wolgemuth2011, Barnhart2011}  is also assumed to be a function of space and time. For simplicity we assume that the bulk viscosity  coefficient $\eta>0$ is constant. The assumption of  infinite compressibility  in (\ref{chapIeq1})  allows  us to decouple the transport of  (actin) density $\rho(x,t)$ from force balance  making the mechanical problem 'statically determinate'. The mass balance equation   $\partial_t\rho+\partial_x(\rho v)=0$    can then be solved independently after the velocity field is determined \cite{Recho2013b,Rechosubm}.  

 We further assume that some  internal mechanism (stiffness of the cell cortex \cite{Boal2002, Sheetz2006, Prost2007, Barnhart2010, Du2012, Loosley2012}, osmotic pressure actively controlled by the channels and pumps on the cell membrane \cite{Jiang2013, Stroka2014}, etc.) maintains a given size $L_0=l_+-l_-$ of the cell. Therefore the stress at the edges must be the same  $\sigma(l_-(t),t)=\sigma(l_+(t),t)=\sigma_0$,  where $\sigma_0(t)$ is then an unknown function.  To model active protrusion we  impose two kinematic Stefan type boundary conditions  characterizing the   rate of actin polymerization $v_{+}>0$ and depolymerization $v_{-}>0$ on the boundaries of the moving segment \cite{Kruse2006, Larripa2006, Julicher2007, Rubinstein2009,Recho2013b}
\begin{equation}\label{chapIkin}
\dot{l}_{\pm}=v_{\pm}+v(l_{\pm}(t),t).
\end{equation}
For consistency,  the overall mass balance must be also respected  on the moving boundaries and  we set 
 $\rho(l_-(t),t)v_-=\rho(l_+(t),t)v_+,$  
which implies an  instantaneous recycling of depolymerized actin from the trailing edge to the leading edge, see \cite{Recho2013,Recho2013b} for more detail on such closure of the treadmilling cycle.  While there is considerable experimental evidence that active polymerization is indeed localized at the leading edge of a crawling cell, the de-polymerization may be spread along the length of the lamellipodium  \cite{Rubinstein2009,Julicher2007}.  However, in the interest of analytic transparency,    such spreading will be ignored in this study, see though \cite{Recho2013b}. 

The  two functions $\chi$ and $\xi$ can be interpreted as infinite dimensional \emph{controls parameters} and found through an optimization procedure.  Even in the absence of a detailed microscopic model governing the rearrangement of these agents  we   still need to subject them  to integral constraints prescribing the average number of adhesion complexes \cite{Barnhart2011} 
\begin{equation}\label{adhtot}
\frac{1}{L_0}\int_{l_-(t)}^{l_+(t)}\xi(x,t)dx=\xi^*,
\end{equation}
where $\xi^*>0$ is a given constant and
\begin{equation}\label{contrtot}
\frac{1}{L_0}\int_{l_-(t)}^{l_+(t)}\chi(x,t)dx=\chi^*,
\end{equation}
where $\chi^*>0$ is another given constant representing the average number of contractile motors \cite{Thoresen2011} . It is clear from (\ref{adhtot},\ref{contrtot}) that since we prescribe the density of active agents, the performance of the self-propulsion machinery will be proportional to the length of the active segment, so the appropriate velocity functional may be also normalized by the total length. 
 
To simplify the analysis  we  further assume that the motion of the active segment is steady \cite{Julicher2007, Rubinstein2009} with unknown velocity $V=\dot{l}_-=\dot{l}_+$  and that  the unknown functions  $\sigma, v$  and the unknown controls $\xi,\chi$ depend exclusively on the appropriately chosen co-moving coordinate $ u=(x-x_0-Vt)/L_0\in[-1/2,1/2]$ .  
Then in dimensionless variables  $\sigma:=\sigma/\chi^*$, $x:=x/\sqrt{\eta/\xi^*}$, $t:=t/(\eta/\chi^*)$,  $\xi:=\xi/\xi^*$ and  $\chi:=\chi/\chi^*$ we obtain the force balance equation
\begin{equation}\label{systtotTW1}
-\frac{1}{L^2}\partial_u\left(\frac{\partial_{u}\sigma(u)}{g_1(u)}\right)+\sigma(u)=g_2(u),
\end{equation}
 where $L :=L_0/\sqrt{\eta/\xi^*}$.  The re-scaled  control functions  
 $$ g_1(u)=\xi(Lu)\geq 0, g_2(u)=\chi(Lu)\geq 0$$  
must satisfy the constraints 
\begin{equation}\label{adcontrtot}
\int_{-1/2}^{1/2}g_1(u)du=\int_{-1/2}^{1/2}g_2(u)du=1.
\end{equation}
The boundary conditions  take the form
\begin{equation}\label{systtotTW}
\left\{ \begin{array}{c}
\sigma (-1/2)=\sigma(1/2)\\
\frac{1}{L^2}\left(\frac{\partial_u\sigma(1/2)}{g_1(1/2)} -\frac{\partial_u\sigma(-1/2)}{g_1(-1/2)}\right)  =- \overline{\Delta V}
\end{array} \right.
\end{equation}
 where
$$\overline{\Delta V}: =\frac{(v_+-v_-)\eta}{\chi^*L_0}.$$ 
The dimensionless velocity of the segment  per length $\overline{V}=V/L$ can be found from the formula
\begin{equation}\label{systtotTW2}
\overline{V}=\overline{V_m}+\frac{1}{2L^2}\left( \frac{\partial_u\sigma(1/2)}{g_1(1/2)}+\frac{\partial_u\sigma(-1/2)}{g_1(-1/2)}\right)
\end{equation}
where 
$$ \overline{V_m}: =\frac{(v_++v_-)\eta}{2\chi^*L_0}.$$ 
Now, if we assume that the two parameters ($\overline{V_m},\overline{\Delta V}$), characterizing actin treadmilling,  are fixed we can  pose the  optimization problem of  finding the controls $g_1(u)$, $g_2(u)$  ensuring the maximization of the normalized velocity $\overline{V}$. This  problem is nontrivial because the  functional $\overline{V} \lbrace g_1,g_2 \rbrace $ is prescribed implicitly through the unknown  solution of the boundary value problem (\ref{systtotTW1},\ref{systtotTW}). To our advantage though this linear elliptic  problem is  classical and is well understood, e.g.  \cite{Mikhlin1960}.   

We observe that parameter $\overline{V_m}$ enters the expression for the  velocity (\ref{systtotTW2}) in an additive way  and does not affect the solution of the optimization problem.  The reason is that   $\overline{V_m}$ characterizes a propulsion mode  associated with simple accretion of the material at the front and  its  simultaneous removal at the rear; when $\overline{V_m}\neq 0$ an a priori polarity is imposed and the problem of motility initiation disappears.  In view of the complete decoupling  of this mode  of self-propulsion from our controls, in what follows we assume without loss of generality that $\overline{V_m}=0$.  

The  second parameter $\overline{\Delta V}$, also characterizing the protrusion strength, does not induce polarity. As we clarify in the next Section, this parameter represents the mechanical action of pushers and the dependence of the crawling velocity  on  $\overline{\Delta V}$, which is now  sensitive to both controls, is  much more subtle than in the case of $\overline{V_m}$.

\section{Pushers and pullers}

 To see that parameter  $\overline{\Delta V}$ characterizes in our setting the activity of 
  pushers,  consider the global balance of couples in the co-moving coordinate system  
\begin{equation}\label{1}
 L \int_{-1/2}^{1/2}g_1(u)v(u)udu=- \overline{\Delta V}+  \int_{-1/2}^{1/2}g_2(u)du+\sigma_0.
\end{equation}
Here the term in the left hand side 
characterizes the total moment due to external (frictional) forces.  The first term in the the right hand side
$$T =\overline{\Delta V}$$
is due to active protrusion, while  the second term
 $ \int_{-1/2}^{1/2}g_2(u)du=1$  
 is due to active contraction.  The last term $\sigma_0$ corresponds to passive reaction forces resulting from the prescription of the  length  of the segment.

Our assumption that  $\overline{\Delta V}>0$ means  that the protrusion  couple  has a negative sign showing that the corresponding force dipoles act on the surrounding medium by pushing outward and creating negative stress. Instead, the  contraction couple  has a positive sign because the contractile  forces pull inward and the induced stresses are positive.  We can therefore associate  protrusion  with the presence of \emph{pushers}  and  contraction with the activity of \emph{pullers} \cite{Carlsson2011, Recho2013}.  We can also (tentatively) argue that motility is protrusion-dominated when $T>1$ and it is contraction-dominated when $0<T<1$.  This assertion will be confirmed later in the paper by rigorous analysis. 

To illustrate the different roles played in  our motility  mechanism by pushers and pullers, we present  below an analysis of a toy model which anticipates the main conclusions of the paper.  We temporarily set $\overline{\Delta V}=0$ and describe the distribution of pushers and pullers by the same function $g_2(u)$ allowing it now to be both positive and negative. Our goal is to show  that for protrusion-dominated motility driven by pushers, it is beneficial to create strong adhesion at the leading edge while for contraction-dominated motility driven by pullers, it is the trailing edge that has to adhere most strongly.

Consider a special   choices of control functions, 
\begin{equation}\label{2}
g_1(u)=q\delta (u-u_1)+(1-q)\delta (u-u_2),
g_2(u)=p\delta (u-u_3)-(1-p)\delta (u-u_4),
\end{equation}
where  $\delta$ is the Dirac distribution, $0\leq q,p\leq 1$ and $-1/2\leq u_1,u_2,u_3,u_4\leq 1/2$.  The control function $g_1(u)$ represents two adhesion sites $u=u_1$ and $u=u_2$ whose locations and intensities  must be optimized  in relation to the prescribed position of  a single puller placed at $u=u_3$ and characterized by a positive dipole moment $p$ and a single pusher located at $u=u_4$ and characterized by a negative dipole moment $p-1$.  The parameter $p$ measures the relative importance of contraction comparing to protrusion. 

Suppose  that
 \begin{equation}\label{22}
u_3<u_1, u_2<u_4
\end{equation}  which ensures that our active segment moves from left to right and that in the segment the adhesion sites are always outside   the  location of the active agents.  By using (\ref{2})   we can express the  velocity of the segment  as
$$\overline{V}=\frac{1}{2}\left( 1+\frac{(1-2p)(u_2-u_1)c}{u_1-u_2+1+(u_2-u_1)a}  \right), $$
where,
$$a=\frac{1}{1+q(1-q)(u_2-u_1)L^2} \text{ and } c=\frac{1-2q}{1+q(1-q)(u_2-u_1)L^2}.$$
Suppose for simplicity that our two adhesive complexes are placed symmetrically with respect to the center of the segment
 $u_1=1-u_2$. We can then use a single parameter  $\Delta u=u_1+1/2=1/2-u_2$ to obtain,
$$\overline{V}=\frac{1}{2}\left(1+\frac{(1-2p)(1-2q)}{1+2\Delta u q(1-q)L^2+\frac{2\Delta u}{1-2\Delta u}}\right).$$
At  $\Delta u=0$, when adhesive complexes localize at the edges of the segment, the velocity reaches its maximum value
\begin{equation}\label{vpushpull}
\overline{V}=\frac{1}{2}\left[ 1+(1-2p)(1-2q)\right] .
\end{equation}
Because of the imposed inequalities (\ref{22}),  we must necessarily have in this configuration $u_3=-1/2$ and $ u_4=1/2$ meaning that pushers and pullers must also localize at the edges. Notice however that inequalities $u_3<u_1$ and $u_2<u_4$ are necessary for formula \eqref{vpushpull} to hold so the apparent conclusion that  $u_3=u_1=1/2$ and $ u_4=u_2=-1/2$ is a result of an abuse of notation. In a theory where, pushers, pullers and adhesion complexes have a characteristic size of dispersion, the adhesion clusters will be slightly ahead of pullers and slightly behind the pushers so that active agents take advantage of the firm attachment to either push or pull (see more on this subject below).

From our simple analysis it follows that  if pullers dominate ($p>1/2$), the optimal way to tune adhesion is to set $q=1$ and concentrate adhesive complexes at the trailing edge of the moving segment achieving maximum velocity  $\overline{V}=1$ when $p=1$.  If, instead,  pushers dominate ($p<1/2$), the optimal way to tune adhesion is to set $q=0$ and concentrate adhesive complexes at the leading edge. In this case, the maximal velocity is again $\overline{V}=1$  given that $p=0$. In other words, to be effective,  pullers have to concentrate on the trailing edge  and ensure strong  adhesion on the leading edge:  in this way pullers can  inflict contraction that displaces the trailing edge which due to the length constraint also propels  the leading edge. On the contrary, pushers can take advantage of the firm attachment  at the trailing edge  to push against it and  propel the leading edge which in turn pulls  the trailing edge due to the length constraint.

\section{Contraction driven motility}\label{Contraction}

We now return to the study of the optimization problem in the original formulation.  The simplest analytically transparent case is when protrusion is disabled $\overline{\Delta V}=0$ and motility is fully contraction-driven. 

Suppose first that  $ g_1\equiv 1$ which means that the  adhesion complexes are distributed uniformly.  
Then the velocity  of the active segment can be expressed as a  quadrature
\begin{equation}\label{contrac}
\overline{V}=-\frac{1}{2\sinh(\frac{L}{2})}\int_{-1/2}^{1/2}\sinh(Lu)g_2(u)du.
\end{equation}
One can see  that if the function $g_2(u)$ is even, then $\overline{V}=0$.  This result can be interpreted as an analogue of the famous Purcell's theorem about the necessity of non-symmetric strokes in Stokes swimming \cite{Purcell1977, Lauga2009}.  If the  distribution $g_2(u)$ is non-symmetric and, for instance,  more motors are placed at the rear of the segment, the  velocity will become positive.  Using the fact $g_2(u)\geq 0$ we can also conclude from  (\ref{contrac})  that  $\overline{V}\leq  1/2.$  This upper bound is reached  when all the motors are fully localized at the rear and  
 $g_2(u)=\delta(u+1/2)$. 

Now, consider the general case when the focal adhesions are distributed inhomogeneously.  Since $\text{(\ref{systtotTW1})} $ is a Sturm-Liouville problem, its solution can be written as
\begin{equation}\label{stessstrum}
\sigma(u)=\sigma_0-\int_{-1/2}^{1/2}G(u,s)\left[g_2(s)-\sigma_0\right]ds, 
\end{equation}
where the Green's function $G(u,s)$ can be represented two  auxiliary functions $h(u)$ and $f(u)$
\begin{equation}\label{stessstrum1}
G(u,s)=\frac{1}{C}\left[ h(u)f(s)\mathbb{1}_{[s<u]}+h(s)f(u)\mathbb{1}_{[u<s]}\right], 
\end{equation}
solving the following standard boundary value problems \cite{Mikhlin1960} :
\begin{equation}\label{hftilde}
\left\{\begin{array}{c}(\frac{1}{g_1}h')'=L^2h\\ h(-1/2)=1,h(1/2)=1\end{array} \right. \text{, }\left\{\begin{array}{c}(\frac{1}{g_1}f')'=L^2f\\ f(-1/2)=1,f(1/2)=-1\end{array} \right. .
\end{equation}
In (\ref{stessstrum1}),   $C=(hf'-fh')/g_1$ is a constant  involving the Wronskian of the two auxiliary functions $h(u)$ and $f(u)$ and $\mathbb{1}$ is the indicator function. 
We can now write
\begin{equation}\label{gencaseadim}
\overline{V}=\frac{1}{2}\int_{-1/2}^{1/2}f(u)(g_2(u)-\hat{g}_2)du,
\end{equation}
where we introduced  a new measure of inhomogeneity of contraction:
$$\hat{g}_2=\frac{\int_{-1/2}^{1/2}h(u)g_2(u)du}{\int_{-1/2}^{1/2}h(u)du}.$$
If both functions  $g_{1,2}(u)$ are even, then $f(u)$ is odd  and,  since the integral of a product of an odd and an even functions  is equal to zero,  we obtain that  $\overline{V}=0$.  The same  result follows if we assume that contraction is   homogeneous   $g_2(u)=\hat{g}_2= 1$  while the adhesion distribution $g_1(u)$ is arbitrary. Therefore, to ensure motility at $\overline{\Delta V}=0$, contraction must be inhomogeneous while  adhesion may still be uniform provided contraction is not even. 
  
To find the optimal distributions  $g_1(u)$, $g_2(u)$ we proceed in two steps. We first show that $\overline{V} \leq 1$ and then find a configuration of controls allowing the cell to reach this bound. 

Notice that we can rewrite $(\ref{gencaseadim})$ in the form 
$$ \overline{V} =\frac{1}{2}\left( \int_{S_+}f(u)(g_2(u)-\hat{g}_2)du+\int_{S_-}f(u)(g_2(u)-\hat{g}_2)du\right) $$
where  we defined the domains $S_-=\left\lbrace u/g_2(u)\leq \hat{g}_2 \right\rbrace $  and  $S_+=\left\lbrace u/g_2(u)>\hat{g}_2 \right\rbrace. $  Applying the maximum principle to (\ref{hftilde}) we obtain that $
1\geq h(u)\geq 0$ and  $h(u)\geq f(u)\geq -h(u)$.
Using the bounds on $f$, we can write  
$$\overline{V} \leq \frac{1}{2}\left( \int_{S_+}h(u)g_2(u)du+\hat{g}_2\int_{S_-}h(u)du \right ).$$
Since the integrands are positive and  $h(u) \leq 1$  it finally follows that
\begin{equation}\label{ineg2}
\overline{V} \leq \int_{-1/2}^{1/2}h(u)g_2(u)du \leq \int_{-1/2}^{1/2}g_2(u)du=1.
\end{equation}
Observe that in the case of a homogeneous distribution of adhesive clusters, the velocity could reach only one half of this maximal value.

We now show that the maximal velocity $\overline{V} =1$ can be reached if both controls $g_1(u)$ and $g_2(u)$ are fully localized.
Take  $\theta>0$ and consider a regularized distribution   
$$g_1(u;\theta)=\frac{1}{\pi}\frac{\theta}{\theta^2+(u-u_1)^2}.$$
For this choice of $g_1(u)$   the auxiliary functions  $h(u)$ and $f(u)$ can be written explicitly in term of Legendre polynomials. In the limit $\theta\rightarrow 0$ and  
 $\lim_{\theta\rightarrow 0}g_1(u;\theta)=\delta(u-u_1)$  we obtain 
 $$h(u)=1\text{ and } f(u)=\left\{ \begin{array}{c}
1 \text{ if } u\leq u_1\\
-1\text{ if } u>u_1.
\end{array} \right. $$
By using these explicit expressions for the auxiliary functions  we can rewrite  $(\text{\ref{gencaseadim}})$ in the form
\begin{equation}\label{Vg2}
\overline{V} =\frac{1}{2}\left[\int_{-1/2}^{u_1}g_2(u)du-\int_{u_1}^{1/2}g_2(u)du-2 u_1 \right].
\end{equation}
if we now suppose that $g_2(u)=\delta(u-u_2)$ the expression for velocity reduces to reduces to
$$\overline{V} =\frac{1}{2}\left\lbrace \begin{array}{c}
1-2u_1\text{ if } u_2<u_1\\
-2u_1\text{ if } u_2=u_1\\
-1-2u_1\text{ if } u_2>u_1
\end{array}\right.$$
It is now clear that the velocity reaches its maximal value  as $u_1\rightarrow-1/2$ while $u_2<u_1$. We can then formally write  $u_2=u_1=-1/2$ and obtain the controls 
 $g_2(u)=g_1(u)=\delta(u+1/2)$  
saturating the bound
 $\overline{V} =1.$  
Notice, however,  that if we assume  directly $u_1=u_2 \rightarrow-1/2$ in (\ref{Vg2}), we obtain  $ \overline{V} =1/2$. This is in agreement with the physical intuition that the anchorage point must be located to the right of the pulling force dipole: in this case the pulling  forces advance the rear  edge of the segment  with minimal slipping.  Mathematically, we encounter here the case of non-commutation of the  limiting procedures  $u_2\rightarrow -1/2$ and  $u_1\rightarrow-1/2$  giving $ \overline{V} =1$ only if the limits are taken in the above order. 

To summarize, the optimization of the distribution of focal adhesions allows the \emph{contraction-dominated} mechanism of cell motility to reach the value of  velocity which is twice as large as when the adhesion is uniform. This means that in order to improve the motility performance  the  adhesion machinery must conspire with the contraction machinery making sure that both the motors and the adhesive centers are localized at the trailing edge. Interestingly, exactly this  type of  correlation between the stresses created by contraction and the distribution of focal adhesions  was observed in  experiments  and numerical simulations \cite{Wang2005, Bershadsky2006, Vicente-Manzanares2009, Wolfenson2009, Shutova2012, Gao2011}.  The  localization of adhesion complexes close to cell edges, where contraction is the strongest, has been also reported outside the motility context \cite{Novak2004, Besser2007, Deshpande2008}.

\section{Upper and lower bounds for velocity in the  general case}\label{generalcase}

We now turn to the general case where both contraction and protrusion are active.  In particular, the protrusive power will be characterized by the parameter  $\overline{\Delta V}=T>0$ which was assumed to be equal to zero in the previous Section. We can then write
\begin{equation}\label{gencasetreadadim}
\overline{V}=\frac{1}{2}\left[  \frac{\int_{-1/2}^{1/2}f(u)du}{\int_{-1/2}^{1/2}h(u)du}T +\int_{-1/2}^{1/2}f(u)(g_2(u)-\hat{g}_2)du\right].
\end{equation}
As we see, the first term in the right hand side  
is  associated with 
  protrusion-based  (or filament-driven \cite{Risler2011})  motility  while,  as we have already seen, the second  term  is the contribution due to contraction-based (or motor-driven \cite{Risler2011} )  motility.  We  notice that  if $g_1(u)$ is even, then $f(u)$ is odd and $h(u)$ is even,  leading to  $$\frac{\int_{-1/2}^{1/2}f(u)du}{\int_{-1/2}^{1/2}h(u)du} =0.$$ If  $g_2(u)$ is also even, then $$\int_{-1/2}^{1/2}f(u)(g_2(u)-\hat{g}_2)du=0.$$ In this case the  velocity of the segment is fully controlled by the   accretion mechanism characterized by the parameter  $\overline{V_m}$. 
 
Consider first the case of protrusion-driven motility by assuming that contraction is homogeneous $g_2(u)\equiv 1$ and therefore does not contribute to the overall velocity.  By using again the 
 maximum principle  we obtain inequalities
$$-1\leq\frac{\int_{-1/2}^{1/2}f(u)du}{\int_{-1/2}^{1/2}h(u)du}\leq 1,$$
leading to the upper bound
\begin{equation}\label{ineg1}
\overline{V}=\overline{V_p}\leq\frac{T}{2}.
\end{equation}
The maximum of the protrusive contribution to velocity is reached  when, $g_1(u)=\delta(u-\frac{1}{2}),$  because in this case  $h =1$  and $f=1$ almost everywhere. Observe, that the optimal solution  in the case of protrusion-driven motility  is in  some sense  \emph{opposite} to the solution  $g_1(u)=\delta(u+1/2)$  obtained in the case of  the contraction-driven motility. 

Based on \eqref{ineg2} and  \eqref{ineg1}  we can now argue that  in the case when  both treadmilling and contraction are present,  an upper bound  for velocity is
$$\overline{V}\leq  \frac{T}{2}+1,$$
however, as we have just seen,  in view of the incompatibility of the corresponding optimal controls, this bound cannot be reached.  The optimal strategy for focal adhesions  would then  involve  a  compromise between  the necessity to localize adhesion at the trailing edge in order to assist the contraction mechanism   and the competing trend to localize adhesion at the leading edge in order to improve the protrusion power of the cell.

\begin{figure}[!h]
\begin{center}
\includegraphics[scale=0.6]{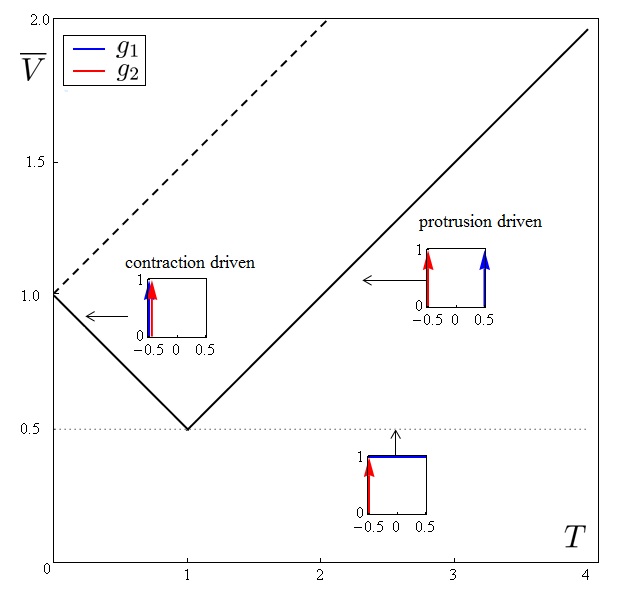}
\caption{\label{contratreadmot} Solid lines: Lower bound on the optimal velocity of self-propulsion $ \overline{V}$ as a function of the measure of the (relative) protrusive strength $T$.  The optimal strategy depends on whether contraction  ($T<1$ ) or protrusion ($T>1$) dominates. The dashed line represents the upper bound obtained by formally summing the  incompatible upper bounds for the protrusion and contraction based strategies. The dotted line represents a sub-optimal  strategy  obtained under the assumption that adhesion is homogeneous.
Insets illustrate the associated configurations of controls $g_1(u)$ and $g_2(u)$.  }
\end{center}
\end{figure}

To obtain a lower bound for $\overline{V}$  we now consider a particular test function representing a weighted sum of our competing  optimal  controls,
 $g_1(u)=q\delta(u+1/2)+(1-q)\delta(u-1/2)$. We  also chose $g_2(u)=\delta(u-u_0),$
where $q\in[0,1]$ and $u_0\in[-1/2,1/2]$ are two parameters to be optimized. Then, by solving (\ref{hftilde})  we obtain,
$$
f(u)=\left\{ \begin{array}{c}
1 \text{ if } u=-1/2\\
\frac{1-2q}{1+q(1-q)L^2}\text{ if } u\in]-1/2,1/2[\\
-1 \text{ if } u=1/2
\end{array} \right.
$$
and,
$$
h(u)=\left\{ \begin{array}{c}
1 \text{ if } u=-1/2\\
\frac{1}{1+q(1-q)L^2}\text{ if } u\in]-1/2,1/2[\\
1 \text{ if } u=1/2, 
\end{array} \right.
$$
which leads to the expression for the velocity
$$\overline{V}= \frac{T}{2}(1-2q)+\frac{1}{2}(f(u_0)-(1-2q)h(u_0)) .$$
The optimization with respect to $u_0$ gives  $u_0=-1/2$ and  
$$\overline{V}=\frac{T}{2}-q(T-1).$$
Finally, optimizing in $q$ we obtain that 
if $T<1$, we need to take $q=0$  and if $T>1$,  we get $q=1$. This result, illustrated  in Fig \ref{contratreadmot}, suggests that 
there is a switch at $T=1$ between the contraction-centered optimization strategy  ($q=0$) and the protrusion-centered  optimization strategy  ($q=1$).  Notice that the switch takes place exactly when the negative protrusion generated  couple $T$ becomes equal to the positive contractile couple equal to $1$. At an interesting state $T=1$, the two active mechanisms neutralize each other and activity-related dipoles become invisible behind the passive contributions in Eq.  (\ref{1}): in this case the optimal position of active and adhesive agents becomes indeterminate.

\section{Numerical solution of the optimization problem}\label{generalcase1}
 
To show that the low bound obtained in the previous Section is  rather close to being optimal,  here we  solve the optimization problem  numerically.  A finite dimensional reduction of the original variational problem is constructed by selecting $N+2$ points 
$u_i= i/(N+1)- 1/2$  that
subdivide the segment [-1/2,1/2]. We then localize adhesion and contraction in these points by choosing the control functions in the form 
$$g_1(u)=\sum_{i=1}^{N}g_1^i\delta(u-u_i) ,  g_2(u)=\sum_{i=0}^{N+1}g_2^i\delta(u-u_i),$$
where $
\sum_{i=0}^{N+1}g_2^i=\sum_{i=1}^{N}g_1^i=1$ and $g_1^i\geq 0,  g_2^i\geq 0.$ In this way we also mimic the  discrete nature of real adhesive clusters and real myosin motors, so in some respects the discrete problem  is even more realistic than the continuum one.

\begin{figure}[!h] 
\begin{center}
\includegraphics[scale=0.4]{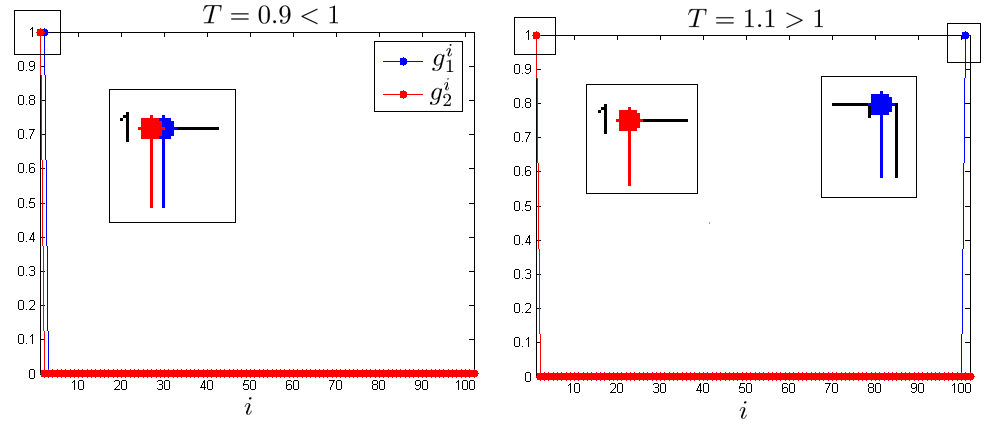}
\caption{\label{loccontraadh}  Numerical results for the  optimization of the function  (\ref{velocitynum}) for two values of the protrusive strength, below the threshold   $T=0.9$ and above the threshold  $T=1.1$. Insets magnify the structure of the optimal controls near the boundaries of the moving segment. Parameter $L=10$.}
\end{center}
\end{figure}

By solving the auxiliary linear elliptic problems for this choice of controls  we obtain that the functions  $h(u)$ and $f(u)$ are piece-wise constant 
 $h(u)=\sum_{i=0}^NA_i\mathbb{1}_{[u_i,u_{i+1}[}(u)$ and $f(u)=\sum_{i=0}^NC_i\mathbb{1}_{[u_i,u_{i+1}[}(u)$. The coefficients with $ i\in[2,N]$ satisfy the equations 
$$g_1^i(A_{i-1}-A_{i-2})+g_1^ig_1^{i-1}A_{i-1}L^2(u_i-u_{i-1})=g_1^{i-1}(A_i-A_{i-1})$$
$$ g_1^i(C_{i-1}-C_{i-2})+g_1^ig_1^{i-1}C_{i-1}L^2(u_i-u_{i-1})=g_1^{i-1}(C_i-C_{i-1})$$
The  boundary conditions give
$$ A_0=1, A_N=1, C_0=1, C_N=-1.$$
We use the conventions $A_{-1}=A_0$, $A_{N+1}=A_N$, $C_{-1}=C_0$ and $C_{N+1}=C_N$ to express the velocity $V$ in the form
\begin{equation}\label{velocitynum}
\overline{V} =\frac{1}{2}\left(T-\sum_{i=0}^{N+1}g_2^i\frac{A_i+A_{i-1}}{2} \right)\frac{\sum_{i=0}^NC_i(u_{i+1}-u_i)}{\sum_{i=0}^NA_i(u_{i+1}-u_i)}+\frac{1}{2}\sum_{i=0}^{N+1}g_2^i\frac{C_i+C_{i-1}}{2}.
\end{equation}
The function  (\ref{velocitynum}) was optimized numerically  with respect to parameters  $ g_1^i $ and $ g_2^i $ subjected to the appropriate  equality and inequality type constraints. To find the global minimum we used  the method of simulated annealing with initial guesses corresponding to homogeneous configurations.  In  Fig.\ref{loccontraadh},  illustrating our results for $N=100$, one can see that for $T<1$ both optimal functions  $g_1(u)$  and $g_2(u)$ are localized at the trailing edge. Instead, for  $T>1$ we observe that  $g_1(u)$ localizes at the leading edge while $g_2(u)$ localizes at the trailing edge. The value of the maximal velocity, obtained numerically, is the same as in the bound up to an error proportional to the mesh size.  Notice also  that  the optimally spaced adhesion points are shifted from the locations of force dipoles by a mesh size.  Thus, the solution of the numerical (regularized) problem  in the contraction-dominated regime (with $T=0.9$), shown  in Fig.~\ref{loccontraadh}, clearly distinguishes the optimal functions $g_1(u)$ and $g_2(u)$ that are both localized at the size of the mesh. The function $g_1(u)$ remains different from zero at one mesh size beyond the point where we already have $g_2(u)=0$ (for positive velocity).  In the  protrusion-dominated regime (at  $T=1.1$) shown in   Fig.~\ref{loccontraadh},  the mesh size again prevents  localization of the function $g_1(u)$  exactly at the leading edge while $g_2(u)$  localizes exactly at the trailing edge. These observations confirm that the regularized numerical solution is more realistic than its singular  analytic prototype. Overall,  our numerical results are in  full agreement with the analytic bounds  which suggests that that these bounds are (nearly) sharp.

\section{Local stability analysis}\label{generalcase11}

In this Section we use a perturbation analysis to provide additional evidence  that our lower bounds are close to being optimal. 

In what follows we use  the superscript   $ \circ$ to indicate the  unperturbed state   and  superscript   $ \star$ to mark parameters characterizing the  perturbed configuration.  We assume that for all functions $\phi(u)$ the following expansion holds in the first order 
$$\phi(u)={\phi}^{\circ}(u)+\epsilon {\phi}^{\star}(u),$$ 
where $\epsilon$ is a  small  parameter.  Keeping only the first order term in  the expression for  $\overline{V}$, we obtain  
$$
\begin{array}{c}
\overline{V}^{\star}=\frac{1}{2}\left[ \int_{-1/2}^{1/2}g_2^{\circ}f^{\star}+f^{\circ}g_2^{\star}-\frac{\int_{-1/2}^{1/2}f^{\circ}}{\int_{-1/2}^{1/2}h^{\circ}}\left(\int_{-1/2}^{1/2}g_2^{\circ}h^{\star}+h^{\circ}g_2^{\star} \right)\right. \\
\left.+\frac{T-\int_{-1/2}^{1/2}h^{\circ}g_2^{\circ}}{\int_{-1/2}^{1/2}h^{\circ}} \left( \int_{-1/2}^{1/2}f^{\star}-\frac{\int_{-1/2}^{1/2}h^{\star}}{\int_{-1/2}^{1/2}h^{\circ}}\int_{-1/2}^{1/2}f^{\circ} \right)   \right], 
\end{array}
$$
where $h^{\star}=H^{\star}{'}$, $f^{\star}=F^{\star}{'} $ and  
\begin{equation}\label{HFpert}
\left\{\begin{array}{c}H^{\star}{''}-L^2g_1^{\circ}H^{\star}=L^2h^{\circ}g_1^{\star}\\ 
H^{\star}{'}(-1/2)=H^{\star}{'}(1/2)=0
\end{array} \right.
\text{, }
\left\{\begin{array}{c}F^{\star}{''}-L^2g_1^{\circ}F^{\star}=L^2F^{\circ}g_1^{\star}\\ 
F^{\star}{'}(-1/2)=F^{\star}{'}(1/2)=0.
\end{array} \right. 
\end{equation}
In view of the saturation of the  constraints by the unperturbed solution
 $\int_{-1/2}^{1/2}g_2^{\circ}=\int_{-1/2}^{1/2}g_1^{\circ}=1,$  we must have 
$$\int_{-1/2}^{1/2}g_2^{\star}=\int_{-1/2}^{1/2}g_1^{\star}=0.$$

Suppose now  that we perturb the 'optimal' controls   $g_1^{\circ}(u)=g_2^{\circ}(u)=\delta(u+ 1/2)$ delivering the lower bound for velocity at  $T<1$.  We can again formally set
$$h^{\circ}(u)\equiv 1\text{ and } f^{\circ}(u)=\left\{ \begin{array}{c}
1 \text{ if } u=-1/2\\
-1\text{ if } u>-1/2,
\end{array} \right. $$
while remembering that the localization point for adhesion must be shifted with respect to the point of localization of  contraction.  
The perturbation of  velocity can be written as
$$\overline{V}^{\star}=\frac{1}{2}\left[\int_{-1/2}^{1/2} (f^{\circ}(u)+ h^{\circ}(u))g_2^{\star}(u)du+(T-1)\int_{-1/2}^{1/2} (f^{\star}(u)+ h^{\star}(u))du \right].$$
A rather general class of  perturbed controls can be represented in the form
\begin{equation} \label{111}
\left\lbrace \begin{array}{c}
g_1^{\star}(u)=-q\delta(u+\frac{1}{2})+ r(u) \text{ with, } \int_{-1/2}^{1/2}r(u)du=q\\
g_2^{\star}(u)=-p\delta(u+\frac{1}{2})+l(u) \text{ with, } \int_{-1/2}^{1/2}l(u)du=p,\\
\end{array}\right. 
\end{equation}
Since  $g_1\geq 0$ and $g_2\geq 0$ we demand that $r(u)\geq 0$ and $l(u)\geq 0$ and therefore also $q\geq 0$ and $p\geq 0$.
We obtain 
\begin{equation} \label{11}
\overline{V}^{\star}=-p+ (T-1)\int_{-1/2}^{1/2}r(u)(u+ 1/2)du.
\end{equation}
From (\ref{11}) we see that if $T\leq 1$, a perturbation of the  controls $g_1^{\circ}(u)=g_2^{\circ}(u)=\delta(u+ 1/2)$ leads to the decrease of the velocity: $\overline{V}^{\star}\leq 0$. Instead, if $T>1$, by choosing $r(u)$ such that,
$$\int_{-1/2}^{1/2}r(u)(u+\frac{1}{2})du>\frac{p}{T-1},$$
we obtain a perturbation  with $\overline{V}^{\star}>0$. These observations  suggest  that beyond the threshold $T=1$ the control function $g_1(u)$ should  be no longer localized at the trailing edge edge.  We also see that the most 'efficient' way to make the velocity  larger  at  $T>1$ is to localize the function $r(u)$ at the leading edge of the segment (at $u=1/2$). 

Consider now a perturbation of  the set of controls $g_1^{\circ}(u)=\delta(u- 1/2)$ and $g_2^{\circ}(u)=\delta(u+ 1/2)$ delivering our lower bound for $T>1$.  With the same abuse of notations as before 
we can write 
$$h^{\circ}(u)\equiv 1\text{ and } f^{\circ}(u)=\left\{ \begin{array}{c}
1 \text{ if } u<1/2\\
-1\text{ if } u=1/2.
\end{array} \right. $$ 
We represent the perturbations in the form
\begin{equation} \label{112}
\left\lbrace \begin{array}{c}
g_1^{\star}(u)=-q\delta(u-\frac{1}{2})+ r(u) \text{ with, } \int_{-1/2}^{1/2}r(u)du=q\geq 0\\
g_2^{\star}(u)=-p\delta(u+\frac{1}{2})+ l(u) \text{ with, } \int_{-1/2}^{1/2}l(u)du=p\geq 0.\
\end{array}\right. 
\end{equation}
where again $r(u)\geq 0$ and $l(u)\geq 0$ and therefore also $q\geq 0$ and $p\geq 0$. The ensuing perturbation of velocity is  
$$\overline{V}^{\star}=\frac{1}{2}\left[\int_{-1/2}^{1/2} (f^{\circ}(u)-h^{\circ}(u))g_2^{\star}(u)du+(T-1)\int_{-1/2}^{1/2} (f^{\star}(u)- h^{\star}(u))du \right] =-(T-1)\int_{-1/2}^{1/2} (\frac{1}{2}-u)r(u)du$$
It is now clear that if $T\geq 1$, then $\overline{V}^{\star}\leq 0$,  showing that the perturbations of  controls  are sub-optimal.
This gives another evidence that the test function providing the lower bound for velocity at $T\geq 1$ are at least close to being optimal.  
 
Based on this analysis we conjecture that the function  $\overline{V}(T)$, representing the optimal velocity, is piece-wise linear with a kink at $T=1$. The presence of a threshold indicates  a  switch from contraction-dominated motility  pattern to protrusion-dominated  motility pattern. As the relative power of protrusion, epitomized by $T$, increases beyond this threshold, the  focal adhesions, maintaining the optimality of the self-propulsion velocity,  must  migrate from the trailing to the leading edge of the active segment. The dynamic migration of adhesion proteins to the edges has been observed in experiments  \cite{Novak2004}. In  real cells, however,   both  edges are usually populated by adhesion complexes and  we can speculate that in this way cells can adjust smoothly to transitions from one driving mode to another.

\section{Discussion}

In this paper we used a simple analytically tractable model of cell motility to study the optimal strategies allowing cells to move faster by actively coordinating spatial distributions of contractile and adhesive agents.   Our study reveals that if adhesion complexes can detect  the dominating mechanism of self propulsion, they can self-organize to ensure  the best performance.

We made specific predictions regarding the advantageous correlations between the  distributions of adhesive and force producing agents and showed that the dependence of the maximal velocity of self-propulsion  on  the relative strength of  contraction and  protrusion may be  non-monotone. In particular, our model predicts that a limited  activation of protrusion will necessarily   lower  the maximal velocity achieved in a purely contractile mode of self-propulsion.  However, as the protrusion strength increases, protrusion can   overtake contraction and the velocity of self-propulsion will    increase beyond the level achieved in the contraction-dominated case. 

Previously we have shown that contraction-driven motility mechanism may be sufficient by itself to explain cell polarization, motility initiation, motility arrest and the symmetrization of a cell before mitosis \cite{Recho2013a, Rechosubm}. From the analysis presented in this paper it becomes evident  that,  if the speed of self propulsion is an issue,  cells should mostly rely on protrusion. More specifically, to maximize its velocity performance after motility initiation a cell must switch from contraction-dominated to protrusion-dominated motility mechanism by increasing the protrusive power and appropriately rearranging the  distribution of adhesive complexes.   It was shown in \cite{Recho2013b} that similar transitions between contraction and protrusion  mechanisms can be used by a cell to accommodate different types of cargos.  
 
To compare our predictions with experiments we can use numerical values of parameters  for  keratocyte fragments   \cite{Kruse2006, Larripa2006, Julicher2007}. We obtain the following rough estimates: $\chi^{*} = 10^3 Pa $, $\xi^{*}= 3 \times 10^{16}Pa \cdot m^{-2} \cdot s$, $ \eta =3\times 10^{4} Pa \cdot s$, $L_0=10   \mu m$, $V_m=8 \mu m \cdot min^{-1} $  and $\Delta V=0.6\mu m \cdot min^{-1}$.  Our first quantitative prediction concerns  the case when  active   treadmilling  of actin is knocked down  and adhesion is homogeneous. As we have shown,  the largest velocity  in this case is reached when all myosin motors are localized at the trailing edge as observed in most eukariotic cells \cite{Verkhovsky1999, Csucs2007, Lombardi2007, Yam2007}.  In dimensional form, the predicted maximal velocity is
 $V =L_0\chi^*/(2\eta)\approx 10\mu m \cdot min^{-1},$ 
which is low  in view of the data on keratocyte fragments suggesting that velocity should be in the interval  $30-40 \mu m \cdot min^{-1}$ \cite{Jilkine2011}.  This is not surprising because  according to our results only half of the total amount of motors is "used" in this case   which implies that adhesion homogeneity is highly sub-optimal. 

If the adhesion inhomogeneity is allowed,  the  configuration   becomes optimal when both myosins and integrins are localized at the trailing edge.  Such  highly  correlated distributions have been observed in  experiments  and  generated in microscale-based numerical models \cite{Novak2004, Bershadsky2006, Besser2007, Deshpande2008, Vicente-Manzanares2009, Wolfenson2009, Shutova2012, Zhang2012}.  In this case, as a result of the  cooperative response,  a cell can be more efficient achieving the  velocity that is two times larger than in the case of homogeneous adhesion:
 $V =L_0\chi^*/\eta \approx 20\mu m \cdot min^{-1}.$  
 
 To be even closer to reality we need to take  active treadmilling into consideration and our estimates suggest that  $T\approx 0.1\ll 1$.  This means that  we are in the  contraction-dominated motility regime for which our velocity bound  gives  more realistic value $V=V_m+L_0\chi^*/\eta\approx 28\mu m \cdot min^{-1} $.  Notice, however, that reducing the value $\chi^*$ by one order of magnitude, which is within the existing error  bounds, we may easily reach the regime where $T>1$ and where it would becomes more beneficial for adhesion clusters to localize at the leading edge conspiring with protrusive elements. A spatial correlation of this type between adhesion and protrusion has been recorded in both  experiments and comprehensive numerical  models, see for instance the data in \cite{Bottino2002, Zajac2008} on nematode spermatozoa. 
 
The proposed model can  be also tested indirectly.  For instance, we know that the location of  adhesion complexes in a moving cell can be  identified by measuring  the distribution of traction forces in the elastic environment \cite{Fournier2010, Peschetola2013}.  If  the adhesive complexes are  found to be shifted towards  the leading edge, we would argue that the cell relies  for its advance mostly on actin treadmilling. If instead the adhesive complexes are preferentially  positioned  at the trailing edge, our model suggests that motility is   mostly driven by contraction. Both predictions can be tested by independent measurements.

An interesting possibility would be if cells could   alternate the location of maximum adhesion  between the trailing edge and  the leading edge in response to oscillations in the level of  activity of pullers and pushers. The evidence of such switching may be that both location are populated with adhesive complexes. We may also recall that the classical mechanism of crawling for eukaryotic cells involves two phases \cite{Abercrombie1980, Bellairs2000,Stossel1993, Alberts2002}: one of them is associated with the creation of protrusions that  push at the front while relying on the stabilization  of the trailing edge and  another  one involves  pulling of the rear (of the cargo) which requires fixation at the front. The switching between these two phases takes place almost periodically  and the associated  reorganization of adhesion clusters from the leading to the trailing edge has been well documented \cite{Ambrosi2009a, Fournier2010}.  To capture such non-steady motility pattern in the framework of our model, the simplifying traveling wave assumption would have to be replaced by a more complex ansatz. 

In conclusion, we emphasize that our interpretations are based on the study of a variational problem whose  analysis revealed some interesting correlations between the spatial arrangement of adhesion and contraction agents and has led us to quantitative  predictions that are in  agreement with experiment.  The prototypical nature of the proposed model, however,  conceals considerable complexity of the actual cell motility phenomenon which involves intricate bio-chemical feedback loops,  geometrically complex mechanical  flows and highly nontrivial rheological behavior. 
In particular, the singular nature of the obtained solutions can be at least partially linked to the fact that treadmilling is modeled schematically, with polymerization and depolymerization processes localized at the edges:  at least one additional control function describing the distribution of pushers is needed to regularize the problem in this respect. The situation is complicated further by the fact that the dominant trade-off condition, controlling the  self-organization of active agents, is still unknown notwithstanding some recent results in this direction \cite{Recho2014}. However, even in the absence of the \emph{definitive} optimization criterion and with minimal assumptions about the inner working of the motility machinery, our study reveals that depending on the task and the available resources a cell may have to modify its mode of operation rather drastically to ensure the best possible performance.

\bigskip

\section{Acknowledgments}
 
The authors are thankful to  F. Alouges, D. Ambrosi, G. Geymonat and A. Zanzottera for helpful discussions. The work of P.R. was supported by Monge Doctoral Fellowship at Ecole Polytechnique.

\section*{\refname}
\bibliography{cellmotility}
\bibliographystyle{vancouver}


\end{document}

%% file: Declarations.tex
\usepackage[latin1]{inputenc}

\usepackage{graphicx}
\usepackage{subfigure}
\usepackage{natbib}
\usepackage{bbold}

\usepackage{bm}
\usepackage{hyperref}

\usepackage{empheq}
\usepackage{xspace}
\usepackage{color}
\usepackage{amssymb}
\usepackage{booktabs}
\usepackage{appendix}

\hypersetup{
bookmarksnumbered = true,
unicode=false,          
pdftoolbar=true,        
pdfmenubar=true,        
pdffitwindow=false,     
pdfstartview={Fit},    
pdftitle={Maximum self-propulsion velocity of  an active segment},    
pdfauthor={P.Recho},     
pdfsubject={Subject},   
pdfcreator={Creator},   
pdfproducer={Producer}, 
pdfkeywords={keyword1} {key2} {key3}, 
pdfnewwindow=true,      
colorlinks=true,       
linkcolor=black,          
citecolor=black,        
filecolor=black,      
urlcolor=black,           
pdfdisplaydoctitle = true
}

\usepackage[normalem]{ulem}
\usepackage{color}
\definecolor{myorange}{rgb}{0.9568,0.4941,0.1961}
\definecolor{myred}{rgb}{0.9098,0.1294,0.2078}
\definecolor{myblue}{rgb}{0.0352,0.4981,0.6509}
\definecolor{mygreen}{rgb}{0.2235,0.6353,0.2588}

\usepackage{marvosym}

